\documentclass{aa}

\usepackage{graphicx}
\usepackage{amsmath,amssymb,amsfonts}
\usepackage{booktabs}
\usepackage{multirow}
\usepackage{xcolor}
\usepackage{placeins}
\usepackage{hyperref}      
\usepackage{lineno}
\let\linenumbers\relax

\begin{document}

   \title{Sub-Snowline Formation of Gas-Giant Planets in Binary Systems}

   \titlerunning{Sub-Snowline Gas Giants in Binaries}

   \author{
     I. Kamai\inst{1}\corrauth{ilay.kamai@campus.technion.ac.il}%
       \thanks{These authors contributed equally to this work.}
     \and H.~B. Perets\inst{1,2}\email{hperets@physics.technion.ac.il}%
       \thanks{These authors contributed equally to this work.}
     \and J. Stegmann\inst{3}
     \and E. Grishin\inst{4,5}
   }

   \authorrunning{Kamai et al.}

   \institute{
     Physics Department, Technion -- Israel Institute of Technology,
       Haifa 32000, Israel
     \and
     ACRO, Open University of Israel, Ra'anana, Israel
     \and
     Max Planck Institute for Astrophysics,
       Karl-Schwarzschild-Str.\ 1, 85748 Garching, Germany
     \and
     School of Physics and Astronomy, Monash University,
       Clayton 3800, Australia
     \and
     OzGrav: ARC Centre of Excellence for Gravitational Wave Discovery,
       Clayton 3800, Australia
   }

   \date{Received \today}


   \abstract
   {Gas-giant planets are thought to require conditions beyond the water
    snow line to build solid cores efficiently. In close binary star
    systems, the companion's gravity additionally limits the region of
    stable orbits, potentially excluding the zone where giants should form.}
   {We aim to identify binary systems in which gas giants exist despite
    the snow line lying in the dynamically unstable zone, and to develop
    a physically motivated formation channel that explains and predicts
    their observed locations.}
   {We analyse a catalogue of 811 circumstellar binary systems from
    \citet{Thebault2025}, identifying those hosting gas giants
    ($M_p \geq 0.15\,M_\mathrm{Jup}$) with snow lines larger than
    $0.8\,a_c$ as defined by \citet{Quarles_2020}. We compare their
    metallicity and eccentricity distributions with the background
    population, model snow-line evolution with MESA, and fit a linear relation between observed planet semi-major axes and the tidal truncation radius from \citet{Pichardo2005}.}
   {Among 393 gas-giant hosts, we identify 17 systems whose snow line lies in the dynamically unstable zone. Their metallicity and eccentricity distributions are consistent with the background population. We propose that a dust trap formed near the tidal truncation radius of the protoplanetary disc can explain sub-snowline giant formation. The observed planet positions follow $a_\mathrm{planet} = (0.569 \pm 0.05)\,r_t$ ($R^2 = 0.94$), enabling  system-by-system predictive power. Evolved systems deviate from this relation, independently supporting a second-generation planet origin for those cases.}
   {The tidal truncation of a protoplanetary disc by the stellar
    companion provides a natural mechanism for sub-snowline gas-giant
    formation in binaries. The resulting empirical relation yields
    testable predictions for binary eccentricities in systems lacking
    direct orbital measurements.}


   \maketitle

\section{Introduction}
\label{sec:intro}

Planetary formation is thought to be a bottom-up process
\cite{Safronov1972}, where dust particles accrete through aerodynamic
and gravitational collisions \citep{Dominik1997, Wada2009}. This
accretion is constrained by the gaseous protoplanetary disc lifetime
(approximately 3--10\,Myr \citep{Erc+17}) and the solids abundances.
For a gas giant to form, it is thought that a solid core exceeding a
critical mass (around 5--10\,$M_\oplus$) should form within the lifetime
of the gaseous disc, allowing a phase of runaway growth and accretion of
its massive gaseous envelope \citep{Greenberg1978, Pollack1996,
Armitage2024, Ikoma2025}. To rapidly reach the critical core mass, dust
aggregation must be highly efficient. An enhanced efficiency can result
from the sharp change in material properties of the disc around the snow
line, the distance to the host star at which the temperature is low
enough for water ice to form ($T_\mathrm{freeze} \approx 170$\,K). Ice
aggregation is approximately ten times more effective compared to typical
materials found below the snow line \citep{Blum2008, Gundlach2015},
suggesting the snow line as a critical lower limit for gas-giant
formation. However, here we identify observational evidence for
sub-snowline formation of gas giants in binaries and develop a
physically motivated mechanism that links measured binary orbits to
predicted planet radii. We further show that the same framework explains
outliers when accounting for stellar evolution or higher multiplicity,
and it yields testable, system-by-system predictions.

When a planet is part of a binary system rather than a single star, the
gravitational tides raised by the stellar companion limit the region
where stable orbits exist. \citet{Holman1999} investigated the stability
of circumstellar orbiting planets in a binary system, and derived the
following relationship for the critical semi-major axis of an inner
(circumstellar) planet:
\begin{equation} \label{eq:stability}
  \frac{a_c}{a_b} = 0.464 - 0.380\mu - 0.631 e_b + 0.586\mu e_b
                    + 0.150 e_b^2 - 0.198 \mu e_b^2.
\end{equation}
Here, $a_c$ is the critical semi-major axis for stability,
$\mu = m_2/(m_1 + m_2)$ is the stellar mass ratio, with $m_2$ being the
companion mass and $m_1$ the primary mass, $e_b$ is the binary
eccentricity, and $a_b$ is the binary semi-major axis. This expression
was derived via least-squares fitting to N-body simulations, and is well
constrained near the centre of the parameter space; accuracy degrades at
extreme mass ratios or eccentricities. It also does not account for
mutual inclination $i$ between the binary and the planet orbit, which
can significantly change the stability boundary. As an improvement,
\citet{Quarles_2020} produced a numerical criterion that accounts for
$a_b, e_b, \mu$, and $i$; we use their tabulated values throughout this
work for all stability estimates.

\begin{figure*}
    \centering
    \includegraphics[width=0.7\textwidth]{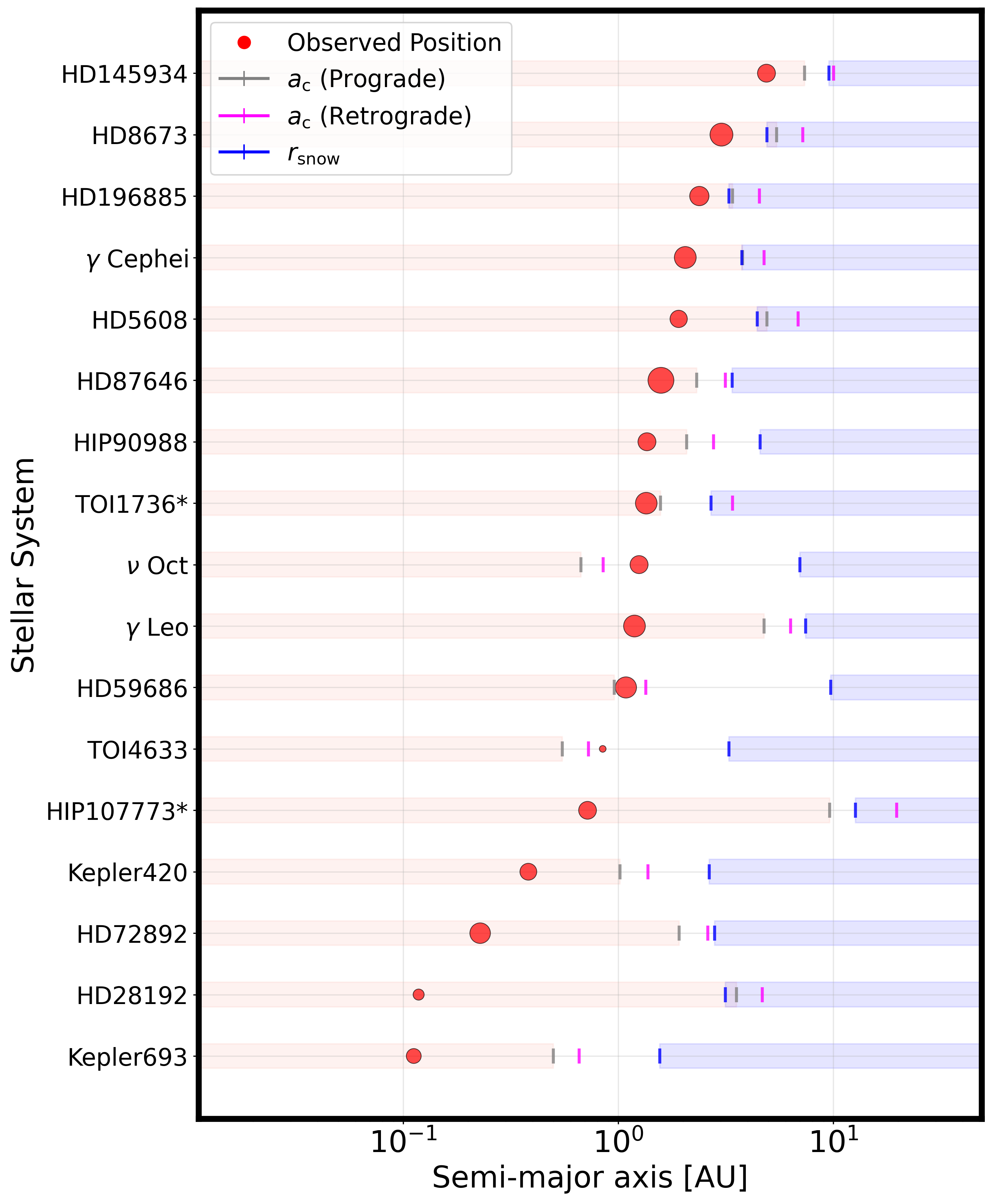}
    \caption{All circumstellar gas giants for which
    $r_\mathrm{snow} > 0.8\,a_c$.
    The x-axis shows the current planetary semi-major axis. $a_c$ and
    $r_\mathrm{snow}$ are marked as grey and blue vertical lines
    (see Sect.~\ref{sec:methods} for calculation details). The pink
    shaded region marks the prograde-stable zone; the blue shaded region
    marks the ice-formation zone. Note that retrograde stability limits
    extend significantly further out (marked by the magenta lines). 
    While retrograde configurations offer a wider stable region \citep{Stegmann2025},
    it is statistically highly improbable that the entirety of our anomalous sample
    consists of retrograde orbiters, retaining the necessity for a distinct 
    formation channel. Systems with an asterisk have no measured
    eccentricity; zero eccentricity is assumed for their stability
    calculation.}
    \label{fig:all_systems}
\end{figure*}

Combining the snow-line and stability constraints, a gas giant in a
circumstellar system faces two simultaneous requirements: the snow line
$r_\mathrm{snow}$ defines a minimum formation distance, while $a_c$
defines a maximum. This leads to a straightforward conclusion: if
$r_\mathrm{snow} > a_c$, gas-giant formation should be impossible.
Surprisingly, we do identify such systems. Analysing a catalogue of 811
circumstellar systems \citep{Thebault2025}, among which 393 host gas
giants, we found 17 gas giants where the snow line is larger than
$0.8\,a_c$ (we choose a conservative buffer of 0.8 to account for the
finite width of both boundaries; see Sect.~\ref{sec:methods}).
Figure~\ref{fig:all_systems} shows all 17 systems with their $a_c$ and
$r_\mathrm{snow}$.

Finding such a sample of planets that does not follow standard theory
has great potential to test our understanding of planet formation
processes, and in particular the formation of gas giants, and
potentially opens the door for new formation channels.

There might be several explanations for this apparent paradox.
In-situ sub-snowline formation of gas giants was suggested to be
potentially possible through (i)~\emph{Pebble accretion at inner
pressure traps} -- at the dead-zone inner edge, silicate sublimation
front, or magnetospheric cavity -- which can concentrate drifting solids
and accelerate core growth; this pathway predicts giants clustered at
relatively small radii, near 0.1--1\,AU for solar-type stars with
metal-enriched envelopes \citep{Lambrechts2012, Lambrechts2014,
Bitsch2015, Chatterjee2014}; (ii)~\emph{Reduced runaway thresholds via
envelope enrichment and low opacities}, which allow gas accretion at
smaller core masses, favouring metal-rich hosts \citep{HoriIkoma2011,
VenturiniHelled2017}; and (iii)~\emph{Local assembly by mergers of
migrating super-Earth cores}, which can push a merged core over the
runaway limit interior to the snow line \citep{HansenMurray2013,
Boley2016}. However, most of our identified planets are at larger
separations than these models suggest (see Fig.~\ref{fig:all_systems}),
and their metallicities are consistent with the background population
(see Fig.~\ref{fig:feh_e_hists}).

Another straightforward possibility is that the snow line evolves with
stellar luminosity and was located in a stable region at the onset of
planet formation. To test this, for each unstable giant we examined the
luminosity of a young single star with the same metallicity and mass,
using Modules for Experiments in Stellar Astrophysics
\citep[MESA;][]{Paxton2011, Jermyn2023}. Out of 13 systems with
measured metallicities and masses, only four had a stable snow line at
formation time (around 2\,Myr; see Fig.~\ref{fig:snowline_evolution}).
While this evolutionary scenario may be physically valid for those four
individual cases, it is insufficient as a general explanation for two
reasons. First, it fails for the remaining systems, whose snow line is
dynamically unstable even at the earliest stellar ages. Second, and more
critically, the snowline-evolution scenario is purely qualitative: it
can address \emph{whether} a giant could have formed, but not
\emph{where} --- it provides no prediction for the specific observed
semi-major axis, which the dust-trap model does.
A further possibility is that the planet formed before the binary
assembled. This scenario is unlikely, as the timescale for binaries to
form at wider separations and migrate inwards is usually much shorter
than the disc lifetime \citep[e.g.][]{Generozov2025}. Capture from a
different system through an exchange interaction would predict a
significant difference in eccentricity between this population and the
general gas-giant population; no such difference is seen
(Fig.~\ref{fig:feh_e_hists}), making this unlikely, though statistics
are small. \citet{Gong2018} investigated the capture of a circumbinary
planet into a close circumstellar orbit, but this probability decreases
dramatically for small binary separations, high eccentricities, or high
mass ratios, and cannot explain the general sample.

Instead, we propose a novel channel for gas-giant planet formation in
binaries, which not only explains their sub-snowline formation but also
predicts the specific system-by-system location of the gas giants.

\section{Methods}
\label{sec:methods}

\subsection{Sample selection}
\label{sec:sample}

We estimate the water-ice snow line for each system in the catalogue.
We assume an irradiated disc where the temperature profile is
\begin{align}
    \frac{T(r)}{T_\odot(1 \, \text{AU})} =
    \left(\frac{L}{L_\odot}\right)^{\frac{1}{4}}
    \left(\frac{r}{1 \, \text{AU}}\right)^{-\frac{1}{2}},
\end{align}
which, when evaluated at the water-ice condensation temperature
($T_\mathrm{freeze} \approx 170$\,K, gives the
snow-line distance as
\begin{align}
    r_{\rm snow} \approx 2.7 \left(\frac{L}{L_\odot}\right)^{\frac{1}{2}}
    \, \text{AU}.
\end{align}
This result is consistent with the positions of gas giants in our solar system. We note that condensation temperature introduces uncertainty that may change $r_\mathrm{snow}$. However, adopting a conservative temperature range of 120--180\,K
\citep{Lodders2003}, does not materially change our
conclusions. Moreover, a more complex disc modelling, which includes viscous heating,
would push the snow line even further out, worsening the sub-snowline formation problem rather
than solving it. Therefore, our simple irradiated disc model provides a safe, conservative
lower limit. We calculated the critical distance for stability, $a_c$, using the
tables from \citet{Quarles_2020}, which provide stability values for
different values of $\mu$, $e_b$, and $i$. We interpolated between
tabulated values and extrapolated for values outside the table boundary
($e_b > 0.8$, for example). We note that the \citet{Quarles_2020}
criterion is based on test-particle N-body simulations and does not
account for the gravitational influence of the gaseous disc on stability
limits or for the finite planetary mass; these represent inherent
limitations that should be borne in mind when interpreting individual
systems. We adopt coplanar ($i = 0$) orbits for systems without measured
inclinations.

We define a gas giant as a planet with $M_p \geq 0.15\,M_\mathrm{Jup}$
(approximately $0.5\,M_\mathrm{Saturn}$) and an `unstable gas giant'
as a gas giant with $r_\mathrm{snow} \geq 0.8\,a_c$.  The factor of 0.8 accounts for the
fact that both $a_c$ and $r_\mathrm{snow}$ represent approximate
boundaries with finite width; resonances have non-zero width and orbital elements undergo secular variations that blur strict cut-offs over time. Varying this conservative buffer to 0.7 or 0.9 yields 19 or
13 systems respectively, without changing the overall results or
conclusions. Out of 811 systems in the catalogue, 393 are gas-giant
hosts. Seventeen are unstable under this criterion, 15 of which have
measured eccentricities.

Taking into account the uncertainty in the binary parameters, one
additional system -- $\tau$~Bootis -- might marginally qualify as
unstable. However, \citet{Justesen2019} showed it likely formed at the
snow line and migrated inward; it is the only system in our sample where
the snow line lies below the truncation radius, making classical
formation feasible. We therefore exclude it from the analysis.

Table~\ref{tab:stellar_params} lists the full sample of 17 unstable
gas-giant systems with their binary and planetary parameters.

\subsection{Truncation radius and the proposed formation mechanism}
\label{sec:mechanism}

As the companion deposits angular momentum into the protoplanetary disc,
the disc is truncated at some radius $r_t$
\citep{Goldreich1979, Goldreich1980, Artymowicz1994, Miranda2015,
Manara2019A&A...628A..95M}. While radial drift is in general faster in
binary systems \citep{Zagaria2021}, a pressure bump can form near $r_t$
from spiral density waves or gas supply from a circumbinary disc
\citep{Poblete2019, Marzari2025}. The formation of a pressure maximum
creates a dust trap, since the drift velocity of dust particles is
proportional to the pressure gradient \citep{Whipple1972,
Weidenschilling1977}. Dust traps have been observed in protoplanetary
discs by ALMA \citep{van_der_Marel2013, Cazzoletti2018, Stadler_2025}
and have been suggested as drivers of planet formation around single
stars. Here we propose they play the same role in binaries.



For the truncation radius, we use the expression of \citet{Pichardo2005},
who derived an analytic fit based on the outermost non-intersecting
invariant loops of test-particle orbits around the primary:
\begin{equation}
r_t = 0.733 \; a_{\rm bin} \mathcal{R}_{\rm RL}(q) (1-e_b)^{1.20}
       \mu^{0.07},
\label{eq:rt}
\end{equation}
where
\begin{equation}
\mathcal{R}_{\rm RL}(q) = \frac{0.49 \, q^{2/3}}{0.6 \, q^{2/3} +
\ln(1+q^{1/3})}
\end{equation}
is the Eggleton Roche-lobe factor \citep{Eggleton1983} and $q = m_1/m_2$ for a disc around the primary star. We prefer this expression over hydrodynamical prescriptions because it is independent
of the viscosity parameter $\alpha$, which is poorly constrained. Sensitivity to the choice of truncation-radius formula is assessed in \ref{app:manara}, where we repeat the analysis using the
observationally calibrated expression of
\citet{Manara2019A&A...628A..95M}.

The dust trap created near $r_t$ also influences migration. For gas
giants, the type-II migration rate follows the radial drift velocity of
the gas due to viscous evolution \citep{Nelson2000}. However, the
elevated dust density in the trap produces feedback on the gas that
reduces the disc torque \citep{Kanagawa2019}, effectively halting
inward migration as long as the trap persists. We therefore assume no
significant migration for this formation channel.

Since both $r_t$ and the dust trap result from the companion--disc
interaction, we hypothesise a correlation between $r_t$ and the final planet position. If the planet forms at the trap and does not migrate
significantly, the observed semi-major axis traces $r_t$ directly. This
can be inverted: if the planet position is known but the binary
eccentricity is not, Eq.~\eqref{eq:rt} constrains the eccentricity at
formation, providing an upper limit since lower eccentricities push
$r_t$ outwards.

\section{Results}
\label{sec:results}

Among the 17 systems in our sample, several warrant careful
consideration before applying the formation channel. \textit{HD~28192}
is part of a triple system \citep{Feng_2022}, and third-body
interactions can significantly alter the disc dynamics.
\textit{Gamma~Leonis}, \textit{HD~59686}, \textit{HIP~90988},
\textit{HD~145934}, and \textit{nu~Octantis} are known to host an
evolved primary or companion \citep{Han2010, Ortiz2016, Jones2021,
Feng2015, cheng_retrograde_2025}: luminosity changes and mass transfer
can have dramatic effects on formation channels, and may instead favour second-generation planet formation from a rejuvenated disc \citep{Perets2010, Stegmann2025}. Crucially, severe stellar mass loss 
adiabatically widens the binary separation and alters the mass ratio $q$, completely erasing the primordial variables needed to preserve the $r_t$ correlation. Two of these -- \textit{nu~Octantis}
and \textit{HD~59686} -- have indeed been suggested to host second-generation planets \citep{cheng_retrograde_2025, Ortiz2016} The remaining three evolved systems are each excluded on independent grounds.
\textit{Gamma~Leonis} hosts two red-clump giants -- both confirmed to have already undergone helium ignition -- with radii and luminosities far exceeding 
their main-sequence values \citep{Takeda2023};
\textit{HD~145934} hosts a K-type giant \citep{Takeda2007} in a compact binary with a low-mass stellar or brown-dwarf companion at only $\sim$22\,AU \citep{Feng2015}; at 
such a small separation, the current orbital architecture almost certainly reflects post-disc dynamical evolution, and the giant host's expansion has erased the primordial mass ratio and binary separation needed to preserve the 
$r_t$ correlation; \textit{HIP~90988} is a Red Giant Branch star whose planet is at $\sim$452 days \citep{Jones2021}. In total, six systems are excluded from the linear-fit analysis for these
reasons; the remaining nine systems with measured eccentricities are
used to calibrate the correlation.

\begin{figure*}
    \centering
    \includegraphics[width=\textwidth]{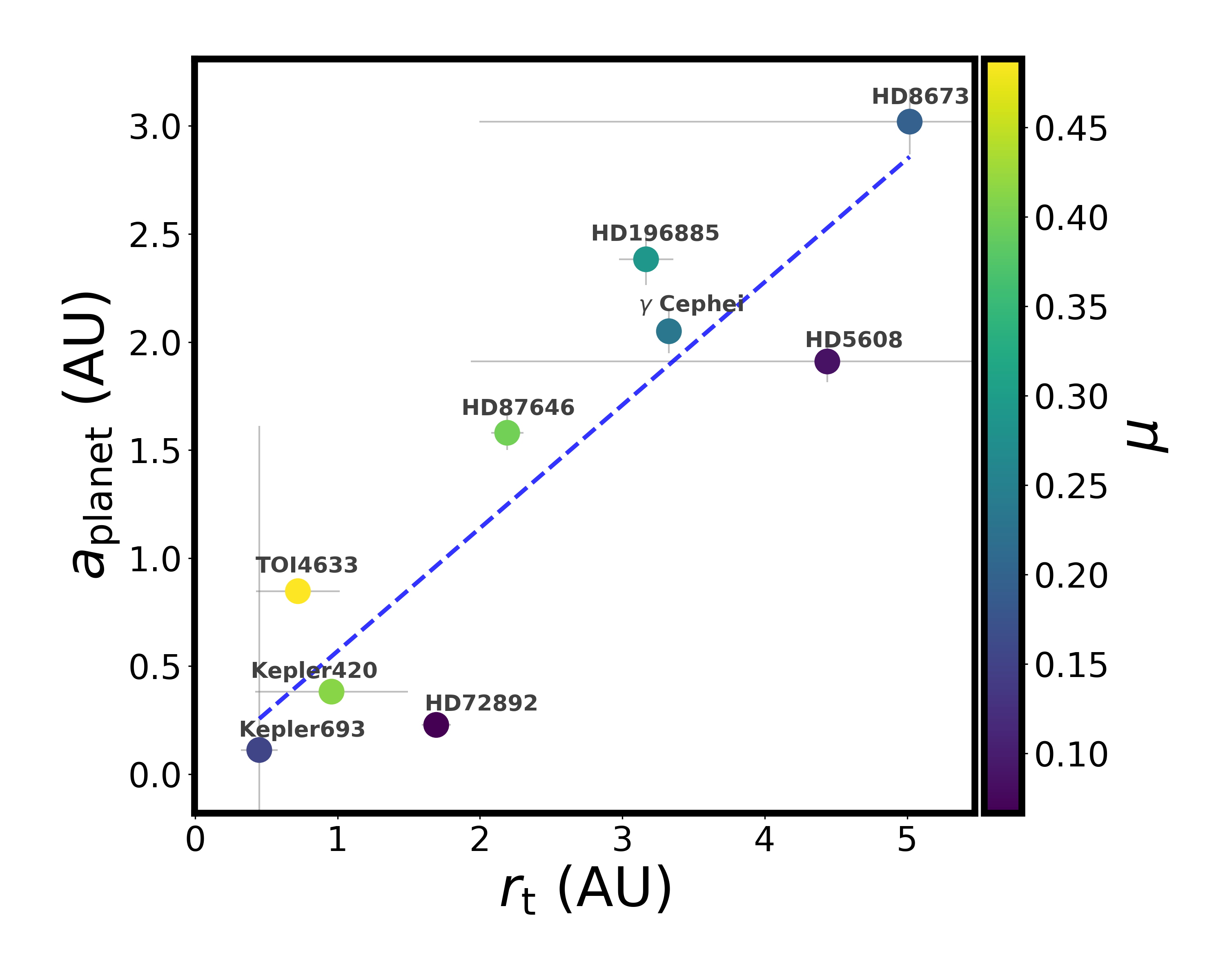}
    \caption{Truncation radius $r_t$ versus observed planetary semi-major
    axis $a_\mathrm{planet}$ for the nine non-evolved, non-triple gas
    giants with measured eccentricity and $r_\mathrm{snow} > 0.8\,a_c$.
    The dashed line shows the best-fit linear relation forced through the
    origin: $a_\mathrm{planet} = 0.569\,r_t$ ($R^2 = 0.94$). Colour
    encodes the companion mass ratio $\mu$. Uncertainties on the x-axis
    are propagated from observed binary parameters; uncertainties on the
    y-axis are taken as 5\% when no uncertainty in $a_\mathrm{planet}$
    was provided.}
    \label{fig:trap_vs_observed_goods}
\end{figure*}

Figure~\ref{fig:trap_vs_observed_goods} shows $r_t$ versus observed $a_\mathrm{planet}$ for these nine systems. A linear fit yields
\begin{equation} a_\mathrm{planet} = (0.569 \pm \mathrm0.05)\, r_t,
    \quad R^2 = 0.94,
\end{equation}
where the fit passes through the origin and the slope uncertainty should be quoted from the fitting procedure. This tight correlation supports
the hypothesis that the dust trap forms at a fixed fraction of the truncation radius, and that subsequent migration is modest.

\begin{figure}
    \centering
    \includegraphics[width=\columnwidth]{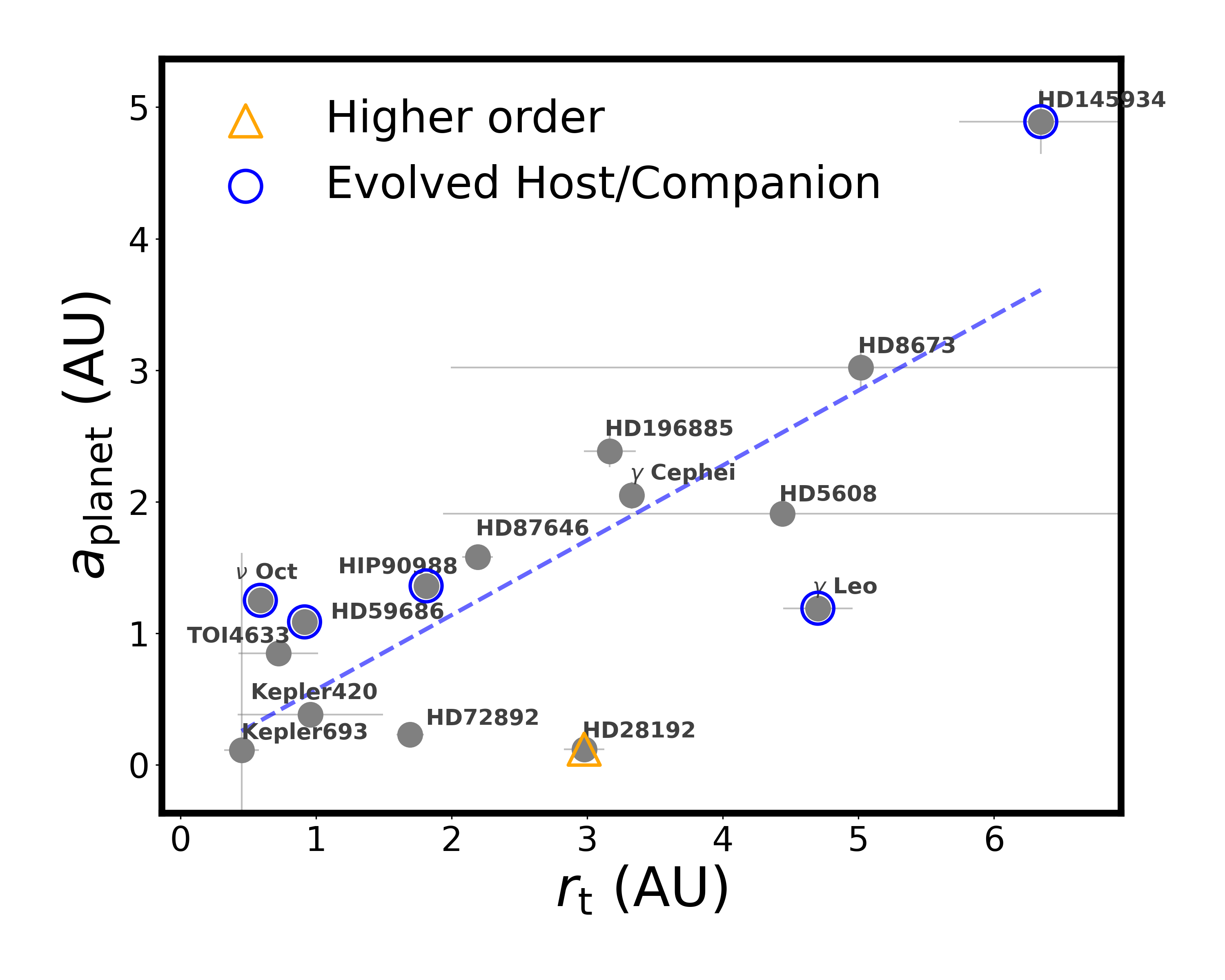}
    \caption{Same as Fig.~\ref{fig:trap_vs_observed_goods} but for the
    full sample. Triple systems are marked with triangles; evolved
    systems with circles. Because evolved and triple systems undergo subsequent dynamical evolution or mass transfer, no linear fit is physically meaningful for them.}
    \label{fig:trap_vs_observed_all}
\end{figure}

Figure~\ref{fig:trap_vs_observed_all} shows
the same relation for the full sample including evolved and triple systems. As expected, evolved systems deviate systematically from the
fitted relation, independently supporting a second-generation planet origin for those cases.

This formation channel also naturally accounts for extreme systems like \textit{TOI~4633}, which was suggested to be in a retrograde orbit by
\citet{Stegmann2025}. Under standard formation theory, the planet would need to form below the snow line; the dust-trap channel provides a
viable alternative. 

Using the same argument but inverting Eq.~\eqref{eq:rt}, we can predict
the binary eccentricity at formation for systems with no measured
eccentricity. Inserting the observed planet semi-major axis as a proxy
for $r_t$ and rearranging:
\begin{equation}
e_b = 1 - \left(\frac{0.569\, a_\mathrm{planet}}
                      {0.733\, a_{\rm bin}\,
                       \mathcal{R}_{\rm RL}(q)\, \mu^{0.07}}
          \right)^{1/1.2}.
\label{eq:e_from_rt}
\end{equation}
This should be interpreted as an upper limit on the eccentricity: if the
binary were more eccentric, $r_t$ would be too small to accommodate the
observed planet position. Adopting a conservative 5\% uncertainty on the
binary semi-major axis, we predict for \textit{HIP~107773} a very high
eccentricity $e_b \lesssim 0.9 \pm 0.01$, and for \textit{TOI~1736}
$e_b \lesssim 0.55 \pm 0.02$. These predictions can be verified by
future astrometric observations; Gaia astrometry provides the most
tractable route, given that the wide projected separations of these
companions make direct radial-velocity determination of the full orbit
challenging. The predictions hold unless the binary eccentricity changed
substantially after planet formation through environmental effects such
as stellar flybys or tidal dissipation.

\subsection{Sensitivity to truncation-radius prescription}
\label{app:manara}
As mentioned in Sect.~\ref{sec:mechanism}, we used Eq.~\eqref{eq:rt}
for the truncation radius due to its simplicity and independence from the viscosity parameter. To assess sensitivity to this choice, we repeated the analysis using the observationally calibrated expression of
\citet{Manara2019A&A...628A..95M}:
\begin{align}\label{eq:rt_manara}
    \frac{r_{t}}{a_b} = R_{\rm RL}\!\left(b\cdot e_b^c + 0.88\,
    \mu^{0.001}\right)
    \cdot \\ \left(\frac{1-e_b^2}{1 + e_b\cos\nu}
    \sqrt{1-\sin^2(\omega + \nu)\sin^2 i}\right)^{-1},
\end{align}
where $b$ and $c$ are parameters fitted to \citet{Artymowicz1994}
results, $\omega$ is the longitude of periastron, $\nu$ is the true anomaly, and $i$ is the inclination. We evaluate at maximum extent ($\omega = 0$, $\nu = 0$). The best-fit slope is $a_\mathrm{planet} = (0.561 \pm 0.06)\,r_t$ with $R^2 = 0.92$, consistent with
the primary result using Eq.~\eqref{eq:rt}, confirming that the choice, the truncation radius prescription does not materially affect our
conclusions.

\begin{figure}
    \centering
    \includegraphics[width=\columnwidth]{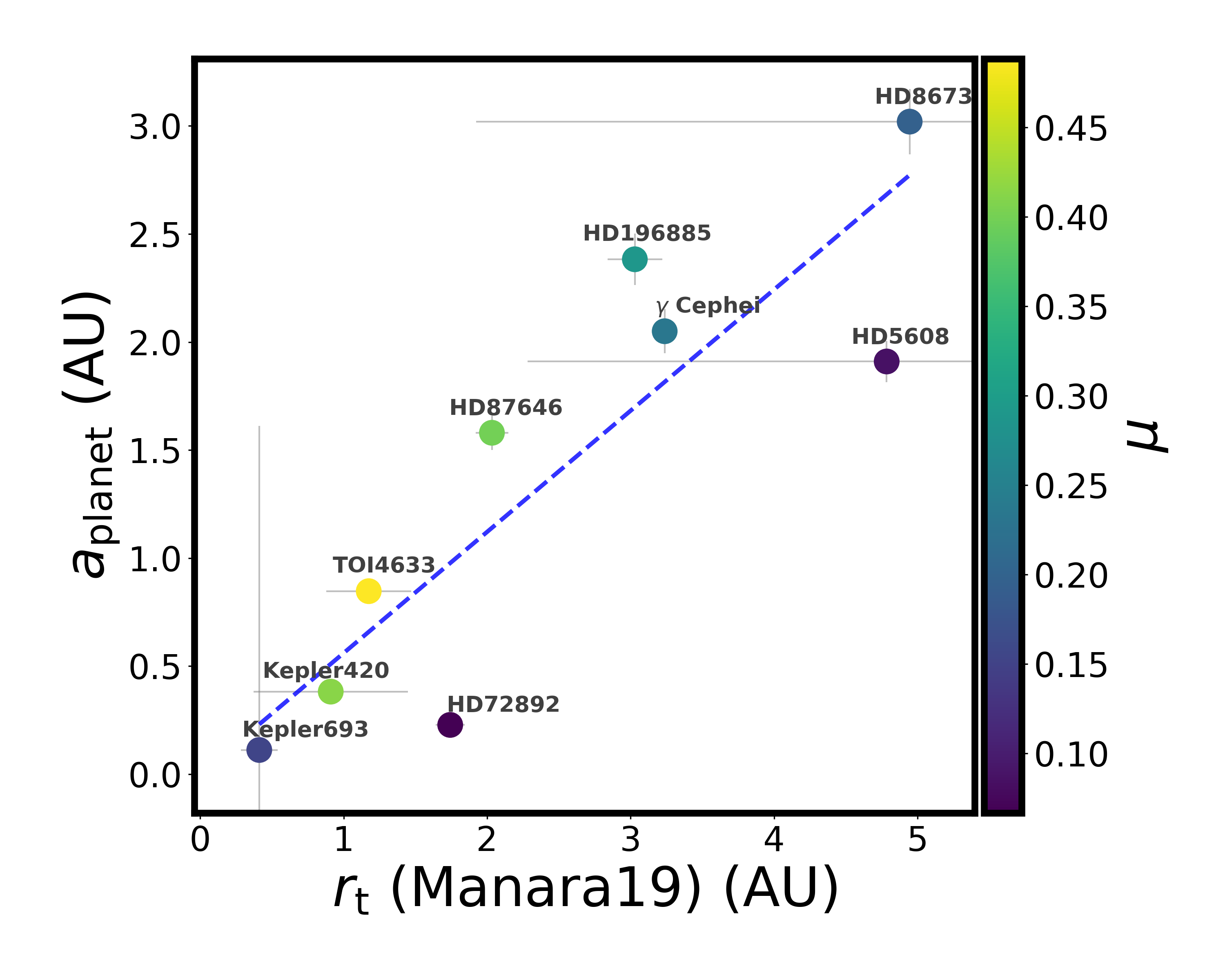}
    \caption{Same as Fig.~\ref{fig:trap_vs_observed_goods} but with
    $r_t$ calculated from \citet{Manara2019A&A...628A..95M}. The dashed
    line is the best-fit linear relation with slope 0.561.}
    \label{fig:rt_manara}
\end{figure}

\section{Discussion}
\label{sec:discussion}

We have shown that gas-giant planets exist in binary systems where
standard formation theory predicts they cannot form, and that their
positions are well described by a linear relation with the tidal
truncation radius. This provides the first system-by-system predictive
framework for gas-giant formation in binary systems.

Planet-formation theory and modelled processes are notoriously difficult
to verify observationally, given the short timescales involved and the
challenges of probing the early stages of disc evolution at high
resolution. The study of snow-line-unstable gas-giant systems in
binaries provides a unique empirical probe into these processes, allowing
quantitative, system-by-system tests of planet formation theory.

We caution that the current sample is small (nine systems used for
calibration), and the linear relation, while tight, rests on
observational data without direct simulation support. The works of
\citet{Poblete2019} and \citet{Marzari2025} lend qualitative theoretical
support to the existence of pressure bumps near $r_t$, but neither
directly simulates the full gas-dust evolution in a circumstellar disc
of the kind studied here. While a full suite of self-consistent hydrodynamical simulations is 
beyond the scope of this primarily observational letter, the tightness of the empirical correlation ($R^2=0.94$)
provides strong motivation for future dedicated theoretical validation of the mechanism.

\section{Conclusions}
\label{sec:conclusions}

We conclude that the sample of gas giants with unstable snow lines must
have formed through a channel distinct from standard core accretion
beyond the snow line. We propose that tidal truncation of the
protoplanetary disc by the stellar companion creates a dust trap at a
fraction $\sim 0.569$ of the truncation radius. The observed planet
positions align well with this prediction ($R^2 = 0.94$), while evolved
systems deviate systematically, further supporting a second-generation
planet origin for those cases.

This model enables, for the first time to our knowledge, system-by-system
predictions of gas-giant planet positions in binary systems. Conversely,
it allows the binary eccentricity (as an upper limit) to be estimated
from an observed planet position, yielding testable predictions for
HIP~107773 ($e_b \lesssim 0.9$) and TOI~1736 ($e_b \lesssim 0.55$).

\begin{acknowledgements}
\end{acknowledgements}

\bibliographystyle{plainnat}   
\bibliography{main}

\begin{thebibliography}{54}
\providecommand{\natexlab}[1]{#1}
\providecommand{\url}[1]{\texttt{#1}}
\expandafter\ifx\csname urlstyle\endcsname\relax
  \providecommand{\doi}[1]{doi: #1}\else
  \providecommand{\doi}{doi: \begingroup \urlstyle{rm}\Url}\fi

\bibitem[{Armitage}(2024)]{Armitage2024}
Philip~J. {Armitage}.
\newblock {Planet formation theory: an overview}.
\newblock \emph{arXiv e-prints}, art. arXiv:2412.11064, December 2024.
\newblock \doi{10.48550/arXiv.2412.11064}.

\bibitem[{Artymowicz} and {Lubow}(1994)]{Artymowicz1994}
Pawel {Artymowicz} and Stephen~H. {Lubow}.
\newblock {Dynamics of Binary-Disk Interaction. I. Resonances and Disk Gap Sizes}.
\newblock \emph{The Astrophysical Journal}, 421:\penalty0 651, February 1994.
\newblock \doi{10.1086/173679}.

\bibitem[{Bitsch} et~al.(2015){Bitsch}, {Lambrechts}, and {Johansen}]{Bitsch2015}
Bertram {Bitsch}, Michiel {Lambrechts}, and Anders {Johansen}.
\newblock {The growth of planets by pebble accretion in evolving protoplanetary discs}.
\newblock \emph{A\&A}, 582:\penalty0 A112, October 2015.
\newblock \doi{10.1051/0004-6361/201526463}.

\bibitem[{Blum} and {Wurm}(2008)]{Blum2008}
J.~{Blum} and G.~{Wurm}.
\newblock {The growth mechanisms of macroscopic bodies in protoplanetary disks.}
\newblock \emph{Annual Review of Astronomy and Astrophysics}, 46:\penalty0 21--56, September 2008.
\newblock \doi{10.1146/annurev.astro.46.060407.145152}.

\bibitem[{Boley} et~al.(2016){Boley}, {Granados Contreras}, and {Gladman}]{Boley2016}
A.~C. {Boley}, A.~P. {Granados Contreras}, and B.~{Gladman}.
\newblock {The In Situ Formation of Giant Planets at Short Orbital Periods}.
\newblock \emph{The Astrophysical Journal Letters}, 817\penalty0 (2):\penalty0 L17, February 2016.
\newblock \doi{10.3847/2041-8205/817/2/L17}.

\bibitem[{Cazzoletti} et~al.(2018){Cazzoletti}, {van Dishoeck}, {Pinilla}, {Tazzari}, {Facchini}, {van der Marel}, {Benisty}, {Garufi}, and {P{\'e}rez}]{Cazzoletti2018}
P.~{Cazzoletti}, E.~F. {van Dishoeck}, P.~{Pinilla}, M.~{Tazzari}, S.~{Facchini}, N.~{van der Marel}, M.~{Benisty}, A.~{Garufi}, and L.~M. {P{\'e}rez}.
\newblock {Evidence for a massive dust-trapping vortex connected to spirals. Multi-wavelength analysis of the HD 135344B protoplanetary disk}.
\newblock \emph{Astronomy \& Astrophysics}, 619:\penalty0 A161, November 2018.
\newblock \doi{10.1051/0004-6361/201834006}.

\bibitem[{Chatterjee} and {Tan}(2014)]{Chatterjee2014}
Sourav {Chatterjee} and Jonathan~C. {Tan}.
\newblock {Inside-out Planet Formation}.
\newblock \emph{The Astrophysical Journal}, 780\penalty0 (1):\penalty0 53, January 2014.
\newblock \doi{10.1088/0004-637X/780/1/53}.

\bibitem[{Cheng} et~al.(2025){Cheng}, {Trifonov}, {Lee}, {Cantalloube}, {Reffert}, {Ramm}, and {Quirrenbach}]{cheng_retrograde_2025}
Ho~Wan {Cheng}, Trifon {Trifonov}, Man~Hoi {Lee}, Faustine {Cantalloube}, Sabine {Reffert}, David {Ramm}, and Andreas {Quirrenbach}.
\newblock {A retrograde planet in a tight binary star system with a white dwarf}.
\newblock \emph{Nature}, 641\penalty0 (8064):\penalty0 866--870, May 2025.
\newblock \doi{10.1038/s41586-025-09006-x}.

\bibitem[{Dominik} and {Tielens}(1997)]{Dominik1997}
C.~{Dominik} and A.~G.~G.~M. {Tielens}.
\newblock {The Physics of Dust Coagulation and the Structure of Dust Aggregates in Space}.
\newblock \emph{The Astrophysical Journal}, 480\penalty0 (2):\penalty0 647--673, May 1997.
\newblock \doi{10.1086/303996}.

\bibitem[{Eggleton}(1983)]{Eggleton1983}
P.~P. {Eggleton}.
\newblock {Aproximations to the radii of Roche lobes.}
\newblock \emph{The Astrophysical Journal}, 268:\penalty0 368--369, May 1983.
\newblock \doi{10.1086/160960}.

\bibitem[{Ercolano} and {Pascucci}(2017)]{Erc+17}
Barbara {Ercolano} and Ilaria {Pascucci}.
\newblock {The dispersal of planet-forming discs: theory confronts observations}.
\newblock \emph{Royal Society Open Science}, 4\penalty0 (4):\penalty0 170114, April 2017.
\newblock \doi{10.1098/rsos.170114}.

\bibitem[Feng et~al.(2022)Feng, Butler, Vogt, Clement, Tinney, Cui, Aizawa, Jones, Bailey, Burt, Carter, Crane, Dotti, Holden, Ma, Ogihara, Oppenheimer, O’Toole, Shectman, Wittenmyer, Wang, Wright, and Xuan]{Feng_2022}
Fabo Feng, R.~Paul Butler, Steven~S. Vogt, Matthew~S. Clement, C.~G. Tinney, Kaiming Cui, Masataka Aizawa, Hugh R.~A. Jones, J.~Bailey, Jennifer Burt, B.~D. Carter, Jeffrey~D. Crane, Francesco~Flammini Dotti, Bradford Holden, Bo~Ma, Masahiro Ogihara, Rebecca Oppenheimer, S.~J. O’Toole, Stephen~A. Shectman, Robert~A. Wittenmyer, Sharon~X. Wang, D.~J. Wright, and Yifan Xuan.
\newblock 3d selection of 167 substellar companions to nearby stars.
\newblock \emph{The Astrophysical Journal Supplement Series}, 262\penalty0 (1):\penalty0 21, aug 2022.
\newblock \doi{10.3847/1538-4365/ac7e57}.
\newblock URL \url{https://dx.doi.org/10.3847/1538-4365/ac7e57}.

\bibitem[{Feng} et~al.(2015){Feng}, {Wright}, {Nelson}, {Wang}, {Ford}, {Marcy}, {Isaacson}, and {Howard}]{Feng2015}
Y.~Katherina {Feng}, Jason~T. {Wright}, Benjamin {Nelson}, Sharon~X. {Wang}, Eric~B. {Ford}, Geoffrey~W. {Marcy}, Howard {Isaacson}, and Andrew~W. {Howard}.
\newblock {The California Planet Survey IV: A Planet Orbiting the Giant Star HD 145934 and Updates to Seven Systems with Long-period Planets}.
\newblock \emph{The Astrophysical Journal}, 800\penalty0 (1):\penalty0 22, February 2015.
\newblock \doi{10.1088/0004-637X/800/1/22}.

\bibitem[{Generozov} et~al.(2025){Generozov}, {Offner}, {Kratter}, {Perets}, {Guszejnov}, and {Grudic}]{Generozov2025}
Aleksey {Generozov}, Stella {Offner}, Kaitlin {Kratter}, Hagai {Perets}, David {Guszejnov}, and Michael {Grudic}.
\newblock {Low mass binaries are bound from birth}.
\newblock In \emph{AAS/Division of Dynamical Astronomy Meeting}, volume~56 of \emph{AAS/Division of Dynamical Astronomy Meeting}, page 501.05, June 2025.

\bibitem[{Goldreich} and {Tremaine}(1979)]{Goldreich1979}
P.~{Goldreich} and S.~{Tremaine}.
\newblock {The excitation of density waves at the Lindblad and corotation resonances by an external potential.}
\newblock \emph{The Astrophysical Journal}, 233:\penalty0 857--871, November 1979.
\newblock \doi{10.1086/157448}.

\bibitem[{Goldreich} and {Tremaine}(1980)]{Goldreich1980}
P.~{Goldreich} and S.~{Tremaine}.
\newblock {Disk-satellite interactions.}
\newblock \emph{The Astrophysical Journal}, 241:\penalty0 425--441, October 1980.
\newblock \doi{10.1086/158356}.

\bibitem[{Gong} and {Ji}(2018)]{Gong2018}
Yan-Xiang {Gong} and Jianghui {Ji}.
\newblock {Formation of S-type planets in close binaries: scattering-induced tidal capture of circumbinary planets}.
\newblock \emph{Monthly Notice of the Royal Astronomical Society}, 478\penalty0 (4):\penalty0 4565--4574, August 2018.
\newblock \doi{10.1093/mnras/sty1300}.

\bibitem[{Greenberg} et~al.(1978){Greenberg}, {Wacker}, {Hartmann}, and {Chapman}]{Greenberg1978}
Richard {Greenberg}, John~F. {Wacker}, William~K. {Hartmann}, and Clark~R. {Chapman}.
\newblock {Planetesimals to planets: Numerical simulation of collisional evolution}.
\newblock \emph{Icarus}, 35\penalty0 (1):\penalty0 1--26, July 1978.
\newblock \doi{10.1016/0019-1035(78)90057-X}.

\bibitem[{Gundlach} and {Blum}(2015)]{Gundlach2015}
B.~{Gundlach} and J.~{Blum}.
\newblock {The Stickiness of Micrometer-sized Water-ice Particles}.
\newblock \emph{The Astrophysical Journal}, 798\penalty0 (1):\penalty0 34, January 2015.
\newblock \doi{10.1088/0004-637X/798/1/34}.

\bibitem[{Han} et~al.(2010){Han}, {Lee}, {Kim}, {Mkrtichian}, {Hatzes}, and {Valyavin}]{Han2010}
Inwoo {Han}, B.~C. {Lee}, K.~M. {Kim}, D.~E. {Mkrtichian}, A.~P. {Hatzes}, and G.~{Valyavin}.
\newblock {Detection of a planetary companion around the giant star {\ensuremath{\gamma}}$^{1}$ Leonis}.
\newblock \emph{Astronomy \& Astrophysics}, 509:\penalty0 A24, January 2010.
\newblock \doi{10.1051/0004-6361/200912536}.

\bibitem[{Hansen} and {Murray}(2013)]{HansenMurray2013}
Brad M.~S. {Hansen} and Norm {Murray}.
\newblock {Testing in Situ Assembly with the Kepler Planet Candidate Sample}.
\newblock \emph{The Astrophysical Journal}, 775\penalty0 (1):\penalty0 53, September 2013.
\newblock \doi{10.1088/0004-637X/775/1/53}.

\bibitem[{Holman} and {Wiegert}(1999)]{Holman1999}
Matthew~J. {Holman} and Paul~A. {Wiegert}.
\newblock {Long-Term Stability of Planets in Binary Systems}.
\newblock \emph{The Astrophysical Journal}, 117\penalty0 (1):\penalty0 621--628, January 1999.
\newblock \doi{10.1086/300695}.

\bibitem[{Hori} and {Ikoma}(2011)]{HoriIkoma2011}
Y.~{Hori} and M.~{Ikoma}.
\newblock {Gas giant formation with small cores triggered by envelope pollution by icy planetesimals}.
\newblock \emph{Monthly Notice of the Royal Astronomical Society}, 416\penalty0 (2):\penalty0 1419--1429, September 2011.
\newblock \doi{10.1111/j.1365-2966.2011.19140.x}.

\bibitem[{Ikoma} and {Kobayashi}(2025)]{Ikoma2025}
Masahiro {Ikoma} and Hiroshi {Kobayashi}.
\newblock {Formation of Giant Planets}.
\newblock \emph{Annual Review of Astronomy and Astrophysics}, 63\penalty0 (1):\penalty0 217--258, August 2025.
\newblock \doi{10.1146/annurev-astro-052722-094843}.

\bibitem[{Jermyn} et~al.(2023){Jermyn}, {Bauer}, {Schwab}, {Farmer}, {Ball}, {Bellinger}, {Dotter}, {Joyce}, {Marchant}, {Mombarg}, {Wolf}, {Sunny Wong}, {Cinquegrana}, {Farrell}, {Smolec}, {Thoul}, {Cantiello}, {Herwig}, {Toloza}, {Bildsten}, {Townsend}, and {Timmes}]{Jermyn2023}
Adam~S. {Jermyn}, Evan~B. {Bauer}, Josiah {Schwab}, R.~{Farmer}, Warrick~H. {Ball}, Earl~P. {Bellinger}, Aaron {Dotter}, Meridith {Joyce}, Pablo {Marchant}, Joey S.~G. {Mombarg}, William~M. {Wolf}, Tin~Long {Sunny Wong}, Giulia~C. {Cinquegrana}, Eoin {Farrell}, R.~{Smolec}, Anne {Thoul}, Matteo {Cantiello}, Falk {Herwig}, Odette {Toloza}, Lars {Bildsten}, Richard H.~D. {Townsend}, and F.~X. {Timmes}.
\newblock {Modules for Experiments in Stellar Astrophysics (MESA): Time-dependent Convection, Energy Conservation, Automatic Differentiation, and Infrastructure}.
\newblock \emph{The Astrophysical Journal}, 265\penalty0 (1):\penalty0 15, March 2023.
\newblock \doi{10.3847/1538-4365/acae8d}.

\bibitem[{Jones} et~al.(2021){Jones}, {Wittenmyer}, {Aguilera-G{\'o}mez}, {Soto}, {Torres}, {Trifonov}, {Jenkins}, {Zapata}, {Sarkis}, {Zakhozhay}, {Brahm}, {Ram{\'\i}rez}, {Santana}, {Vines}, {D{\'\i}az}, {Vu{\v{c}}kovi{\'c}}, and {Pantoja}]{Jones2021}
M.~I. {Jones}, R.~{Wittenmyer}, C.~{Aguilera-G{\'o}mez}, M.~G. {Soto}, P.~{Torres}, T.~{Trifonov}, J.~S. {Jenkins}, A.~{Zapata}, P.~{Sarkis}, O.~{Zakhozhay}, R.~{Brahm}, R.~{Ram{\'\i}rez}, F.~{Santana}, J.~I. {Vines}, M.~R. {D{\'\i}az}, M.~{Vu{\v{c}}kovi{\'c}}, and B.~{Pantoja}.
\newblock {Four Jovian planets around low-luminosity giant stars observed by the EXPRESS and PPPS}.
\newblock \emph{A\&A}, 646:\penalty0 A131, February 2021.
\newblock \doi{10.1051/0004-6361/202038555}.

\bibitem[{Justesen} and {Albrecht}(2019)]{Justesen2019}
A.~B. {Justesen} and S.~{Albrecht}.
\newblock {Constraining the orbit of the planet-hosting binary {\ensuremath{\tau}} Bo{\"o}tis. Clues about planetary formation and migration}.
\newblock \emph{Astronomy \& Astrophysics}, 625:\penalty0 A59, May 2019.
\newblock \doi{10.1051/0004-6361/201834368}.

\bibitem[{Kanagawa}(2019)]{Kanagawa2019}
Kazuhiro~D. {Kanagawa}.
\newblock {Termination of Inward Migration for a Gap-opening Planet through Dust Feedback}.
\newblock \emph{The Astrophysical Journal letter}, 879\penalty0 (2):\penalty0 L19, July 2019.
\newblock \doi{10.3847/2041-8213/ab2a0f}.

\bibitem[{Lambrechts} and {Johansen}(2012)]{Lambrechts2012}
M.~{Lambrechts} and A.~{Johansen}.
\newblock {Rapid growth of gas-giant cores by pebble accretion}.
\newblock \emph{A\&A}, 544:\penalty0 A32, August 2012.
\newblock \doi{10.1051/0004-6361/201219127}.

\bibitem[{Lambrechts} and {Johansen}(2014)]{Lambrechts2014}
M.~{Lambrechts} and A.~{Johansen}.
\newblock {Forming the cores of giant planets from the radial pebble flux in protoplanetary discs}.
\newblock \emph{A\&A}, 572:\penalty0 A107, December 2014.
\newblock \doi{10.1051/0004-6361/201424343}.

\bibitem[{Lodders}(2003)]{Lodders2003}
Katharina {Lodders}.
\newblock {Solar System Abundances and Condensation Temperatures of the Elements}.
\newblock \emph{\apj}, 591\penalty0 (2):\penalty0 1220--1247, July 2003.
\newblock \doi{10.1086/375492}.

\bibitem[{Manara} et~al.(2019){Manara}, {Tazzari}, {Long}, {Herczeg}, {Lodato}, {Rota}, {Cazzoletti}, {van der Plas}, {Pinilla}, {Dipierro}, {Edwards}, {Harsono}, {Johnstone}, {Liu}, {Menard}, {Nisini}, {Ragusa}, {Boehler}, and {Cabrit}]{Manara2019A&A...628A..95M}
C.~F. {Manara}, M.~{Tazzari}, F.~{Long}, G.~J. {Herczeg}, G.~{Lodato}, A.~A. {Rota}, P.~{Cazzoletti}, G.~{van der Plas}, P.~{Pinilla}, G.~{Dipierro}, S.~{Edwards}, D.~{Harsono}, D.~{Johnstone}, Y.~{Liu}, F.~{Menard}, B.~{Nisini}, E.~{Ragusa}, Y.~{Boehler}, and S.~{Cabrit}.
\newblock {Observational constraints on dust disk sizes in tidally truncated protoplanetary disks in multiple systems in the Taurus region}.
\newblock \emph{Astronomy \& Astrophysics}, 628:\penalty0 A95, August 2019.
\newblock \doi{10.1051/0004-6361/201935964}.

\bibitem[{Marzari} and {D'Angelo}(2025)]{Marzari2025}
Francesco {Marzari} and Gennaro {D'Angelo}.
\newblock {Dust supply to close binary systems}.
\newblock \emph{Astronomy \& Astrophysics}, 695:\penalty0 A53, March 2025.
\newblock \doi{10.1051/0004-6361/202453035}.

\bibitem[{Miranda} and {Lai}(2015)]{Miranda2015}
Ryan {Miranda} and Dong {Lai}.
\newblock {Tidal truncation of inclined circumstellar and circumbinary discs in young stellar binaries}.
\newblock \emph{Monthly Notice of the Royal Astronomical Society}, 452\penalty0 (3):\penalty0 2396--2409, September 2015.
\newblock \doi{10.1093/mnras/stv1450}.

\bibitem[{Nelson} et~al.(2000){Nelson}, {Papaloizou}, {Masset}, and {Kley}]{Nelson2000}
Richard~P. {Nelson}, John C.~B. {Papaloizou}, Fr{\'e}d{\'e}ric {Masset}, and Willy {Kley}.
\newblock {The migration and growth of protoplanets in protostellar discs}.
\newblock \emph{Monthly Notice of the Royal Astronomical Society}, 318\penalty0 (1):\penalty0 18--36, October 2000.
\newblock \doi{10.1046/j.1365-8711.2000.03605.x}.

\bibitem[{Ortiz} et~al.(2016){Ortiz}, {Reffert}, {Trifonov}, {Quirrenbach}, {Mitchell}, {Nowak}, {Buenzli}, {Zimmerman}, {Bonnefoy}, {Skemer}, {Defr{\`e}re}, {Lee}, {Fischer}, and {Hinz}]{Ortiz2016}
Mauricio {Ortiz}, Sabine {Reffert}, Trifon {Trifonov}, Andreas {Quirrenbach}, David~S. {Mitchell}, Grzegorz {Nowak}, Esther {Buenzli}, Neil {Zimmerman}, Micka{\"e}l {Bonnefoy}, Andy {Skemer}, Denis {Defr{\`e}re}, Man~Hoi {Lee}, Debra~A. {Fischer}, and Philip~M. {Hinz}.
\newblock {Precise radial velocities of giant stars. IX. HD 59686 Ab: a massive circumstellar planet orbiting a giant star in a 13.6 au eccentric binary system}.
\newblock \emph{A\&A}, 595:\penalty0 A55, October 2016.
\newblock \doi{10.1051/0004-6361/201628791}.

\bibitem[{Paxton} et~al.(2011){Paxton}, {Bildsten}, {Dotter}, {Herwig}, {Lesaffre}, and {Timmes}]{Paxton2011}
B.~{Paxton}, L.~{Bildsten}, A.~{Dotter}, F.~{Herwig}, P.~{Lesaffre}, and F.~{Timmes}.
\newblock {Modules for Experiments in Stellar Astrophysics (MESA)}.
\newblock \emph{The Astrophysical Journal}, 192:\penalty0 3, jan 2011.
\newblock \doi{10.1088/0067-0049/192/1/3}.

\bibitem[{Perets}(2010)]{Perets2010}
Hagai~B. {Perets}.
\newblock {Second generation planets}.
\newblock \emph{arXiv e-prints}, art. arXiv:1001.0581, January 2010.
\newblock \doi{10.48550/arXiv.1001.0581}.

\bibitem[{Pichardo} et~al.(2005){Pichardo}, {Sparke}, and {Aguilar}]{Pichardo2005}
Barbara {Pichardo}, Linda~S. {Sparke}, and Luis~A. {Aguilar}.
\newblock {Circumstellar and circumbinary discs in eccentric stellar binaries}.
\newblock \emph{Monthly Notice of the Royal Astronomical Society}, 359\penalty0 (2):\penalty0 521--530, May 2005.
\newblock \doi{10.1111/j.1365-2966.2005.08905.x}.

\bibitem[{Poblete} et~al.(2019){Poblete}, {Cuello}, and {Cuadra}]{Poblete2019}
Pedro~P. {Poblete}, Nicol{\'a}s {Cuello}, and Jorge {Cuadra}.
\newblock {Dusty clumps in circumbinary discs}.
\newblock \emph{Monthly Notice of the Royal Astronomical Society}, 489\penalty0 (2):\penalty0 2204--2215, October 2019.
\newblock \doi{10.1093/mnras/stz2297}.

\bibitem[{Pollack} et~al.(1996){Pollack}, {Hubickyj}, {Bodenheimer}, {Lissauer}, {Podolak}, and {Greenzweig}]{Pollack1996}
James~B. {Pollack}, Olenka {Hubickyj}, Peter {Bodenheimer}, Jack~J. {Lissauer}, Morris {Podolak}, and Yuval {Greenzweig}.
\newblock {Formation of the Giant Planets by Concurrent Accretion of Solids and Gas}.
\newblock \emph{Icarus}, 124\penalty0 (1):\penalty0 62--85, November 1996.
\newblock \doi{10.1006/icar.1996.0190}.

\bibitem[Quarles et~al.(2020)Quarles, Li, Kostov, and Haghighipour]{Quarles_2020}
Billy Quarles, Gongjie Li, Veselin Kostov, and Nader Haghighipour.
\newblock Orbital stability of circumstellar planets in binary systems.
\newblock \emph{The Astronomical Journal}, 159\penalty0 (3):\penalty0 80, feb 2020.
\newblock \doi{10.3847/1538-3881/ab64fa}.
\newblock URL \url{https://doi.org/10.3847/1538-3881/ab64fa}.

\bibitem[{Safronov}(1972)]{Safronov1972}
V.~S. {Safronov}.
\newblock \emph{{Evolution of the protoplanetary cloud and formation of the earth and planets}}.
\newblock 1972.

\bibitem[Stadler et~al.(2025)Stadler, Benisty, Winter, Izquierdo, Longarini, Galloway-Sprietsma, Curone, Andrews, Bae, Facchini, Rosotti, Teague, Barraza-Alfaro, Cataldi, Cuello, Czekala, Fasano, Flock, Fukagawa, Garg, Hall, Hammond, Hilder, Huang, Ilee, Kanagawa, Lesur, Lodato, Loomis, Menard, Orihara, Pinte, Price, Yen, Wafflard-Fernandez, Wilner, Wölfer, Yoshida, and Zawadzki]{Stadler_2025}
Jochen Stadler, Myriam Benisty, Andrew~J. Winter, Andrés~F. Izquierdo, Cristiano Longarini, Maria Galloway-Sprietsma, Pietro Curone, Sean~M. Andrews, Jaehan Bae, Stefano Facchini, Giovanni Rosotti, Richard Teague, Marcelo Barraza-Alfaro, Gianni Cataldi, Nicolás Cuello, Ian Czekala, Daniele Fasano, Mario Flock, Misato Fukagawa, Himanshi Garg, Cassandra Hall, Iain Hammond, Thomas Hilder, Jane Huang, John~D. Ilee, Kazuhiro Kanagawa, Geoffroy Lesur, Giuseppe Lodato, Ryan~A. Loomis, Francois Menard, Ryuta Orihara, Christophe Pinte, Daniel~J. Price, Hsi-Wei Yen, Gaylor Wafflard-Fernandez, David~J. Wilner, Lisa Wölfer, Tomohiro~C. Yoshida, and Brianna Zawadzki.
\newblock exoalma. vi. rotating under pressure: Rotation curves, azimuthal velocity substructures, and gas pressure variations.
\newblock \emph{The Astrophysical Journal Letters}, 984\penalty0 (1):\penalty0 L11, apr 2025.
\newblock \doi{10.3847/2041-8213/adb152}.
\newblock URL \url{https://doi.org/10.3847/2041-8213/adb152}.

\bibitem[{Stegmann} et~al.(2025){Stegmann}, {Grishin}, {Johnston}, {Eisner}, {Justham}, {de Mink}, and {Perets}]{Stegmann2025}
Jakob {Stegmann}, Evgeni {Grishin}, Cole {Johnston}, Nora~L. {Eisner}, Stephen {Justham}, Selma~E. {de Mink}, and Hagai~B. {Perets}.
\newblock {Planet formation and long-term stability in a very eccentric stellar binary}.
\newblock \emph{arXiv e-prints}, art. arXiv:2501.05506, January 2025.
\newblock \doi{10.48550/arXiv.2501.05506}.

\bibitem[{Takeda} et~al.(2007){Takeda}, {Ford}, {Sills}, {Rasio}, {Fischer}, and {Valenti}]{Takeda2007}
Genya {Takeda}, Eric~B. {Ford}, Alison {Sills}, Frederic~A. {Rasio}, Debra~A. {Fischer}, and Jeff~A. {Valenti}.
\newblock {Structure and Evolution of Nearby Stars with Planets. II. Physical Properties of \raisebox{-0.5ex}\textasciitilde1000 Cool Stars from the SPOCS Catalog}.
\newblock \emph{\apjs}, 168\penalty0 (2):\penalty0 297--318, February 2007.
\newblock \doi{10.1086/509763}.

\bibitem[{Takeda}(2023)]{Takeda2023}
Yoichi {Takeda}.
\newblock {Spectroscopic comparative study of the red giant binary system gamma Leonis A and B}.
\newblock \emph{Astrophysics and Space Science}, 368\penalty0 (7):\penalty0 56, July 2023.
\newblock \doi{10.1007/s10509-023-04214-1}.

\bibitem[{Thebault} and {Bonanni}(2025)]{Thebault2025}
Philippe {Thebault} and Danilo {Bonanni}.
\newblock {A complete census of planet-hosting binaries}.
\newblock \emph{arXiv e-prints}, art. arXiv:2506.18759, June 2025.
\newblock \doi{10.48550/arXiv.2506.18759}.

\bibitem[{van der Marel} et~al.(2013){van der Marel}, {van Dishoeck}, {Bruderer}, {Birnstiel}, {Pinilla}, {Dullemond}, {van Kempen}, {Schmalzl}, {Brown}, {Herczeg}, {Mathews}, and {Geers}]{van_der_Marel2013}
Nienke {van der Marel}, Ewine~F. {van Dishoeck}, Simon {Bruderer}, Til {Birnstiel}, Paola {Pinilla}, Cornelis~P. {Dullemond}, Tim~A. {van Kempen}, Markus {Schmalzl}, Joanna~M. {Brown}, Gregory~J. {Herczeg}, Geoffrey~S. {Mathews}, and Vincent {Geers}.
\newblock {A Major Asymmetric Dust Trap in a Transition Disk}.
\newblock \emph{Science}, 340\penalty0 (6137):\penalty0 1199--1202, June 2013.
\newblock \doi{10.1126/science.1236770}.

\bibitem[{Venturini} and {Helled}(2017)]{VenturiniHelled2017}
Julia {Venturini} and Ravit {Helled}.
\newblock {The Formation of Mini-Neptunes}.
\newblock \emph{The Astrophysical Journal}, 848\penalty0 (2):\penalty0 95, October 2017.
\newblock \doi{10.3847/1538-4357/aa8cd0}.

\bibitem[{Wada} et~al.(2009){Wada}, {Tanaka}, {Suyama}, {Kimura}, and {Yamamoto}]{Wada2009}
Koji {Wada}, Hidekazu {Tanaka}, Toru {Suyama}, Hiroshi {Kimura}, and Tetsuo {Yamamoto}.
\newblock {Collisional Growth Conditions for Dust Aggregates}.
\newblock \emph{The Astrophysical Journal}, 702\penalty0 (2):\penalty0 1490--1501, September 2009.
\newblock \doi{10.1088/0004-637X/702/2/1490}.

\bibitem[{Weidenschilling}(1977)]{Weidenschilling1977}
S.~J. {Weidenschilling}.
\newblock {Aerodynamics of solid bodies in the solar nebula.}
\newblock \emph{Monthly Notice of the Royal Astronomical Society}, 180:\penalty0 57--70, July 1977.
\newblock \doi{10.1093/mnras/180.2.57}.

\bibitem[{Whipple}(1972)]{Whipple1972}
F.~L. {Whipple}.
\newblock {On certain aerodynamic processes for asteroids and comets}.
\newblock In Aina {Elvius}, editor, \emph{From Plasma to Planet}, page 211, January 1972.

\bibitem[{Zagaria} et~al.(2021){Zagaria}, {Rosotti}, and {Lodato}]{Zagaria2021}
Francesco {Zagaria}, Giovanni~P. {Rosotti}, and Giuseppe {Lodato}.
\newblock {On dust evolution in planet-forming discs in binary systems - I. Theoretical and numerical modelling: radial drift is faster in binary discs}.
\newblock \emph{Monthly Notice of the Royal Astronomical Society}, 504\penalty0 (2):\penalty0 2235--2252, June 2021.
\newblock \doi{10.1093/mnras/stab985}.

\end{thebibliography}

\onecolumn
\begin{appendix}

\section{Supplementary Figures}\label{app:figures}

\begin{figure}[h]
    \centering
    \includegraphics[width=\columnwidth]{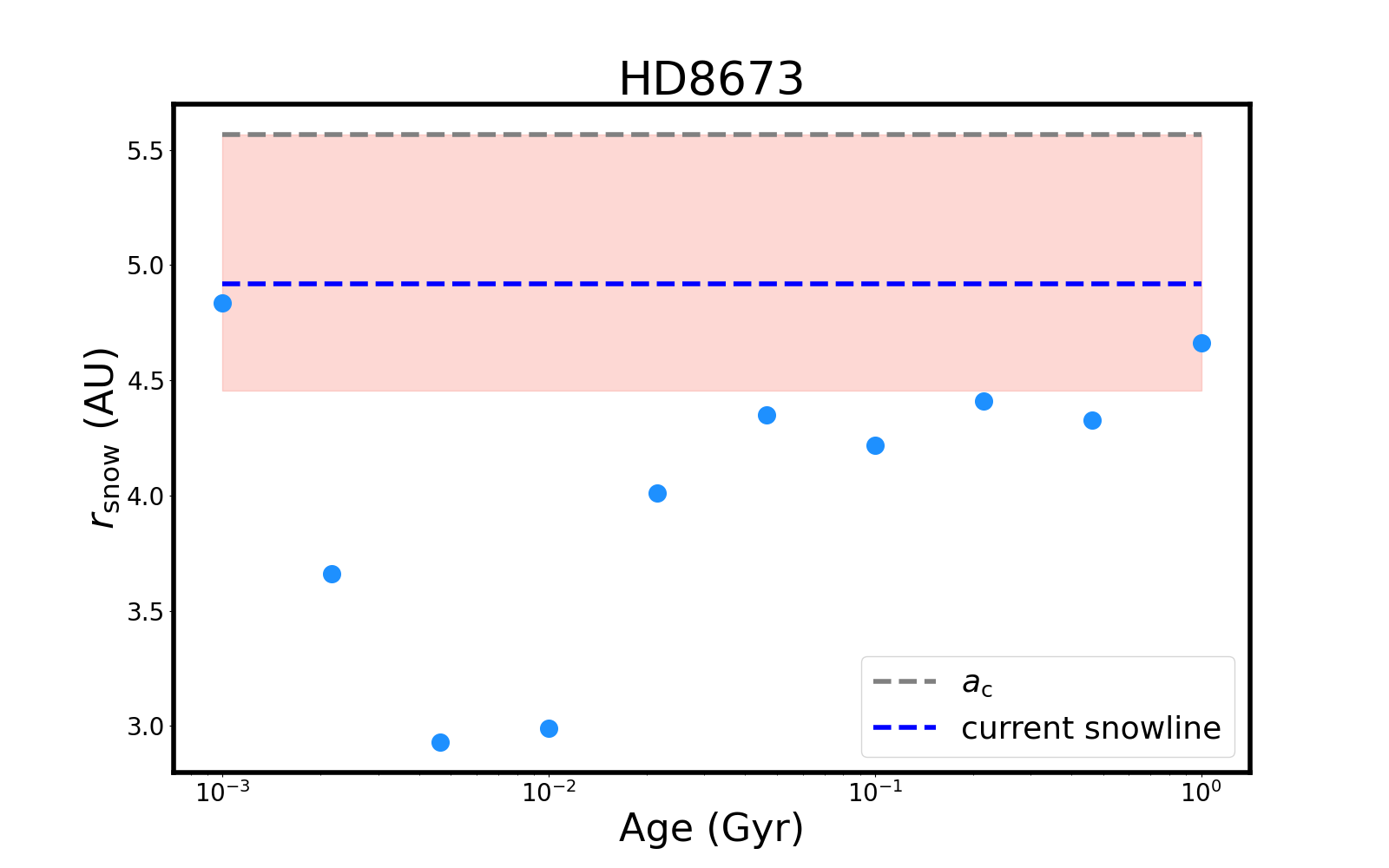}
    \caption{Example of snow-line evolution for \textit{HD~8673}. The
    position of the snow line at different ages was calculated using MESA
    \citep{Paxton2011, Jermyn2023}. The shaded pink region marks the
    unstable zone ($r_\mathrm{snow} > 0.8\,a_c$).}
    \label{fig:snowline_evolution}
\end{figure}

\begin{figure}[h]
    \centering
    \includegraphics[width=0.48\textwidth]{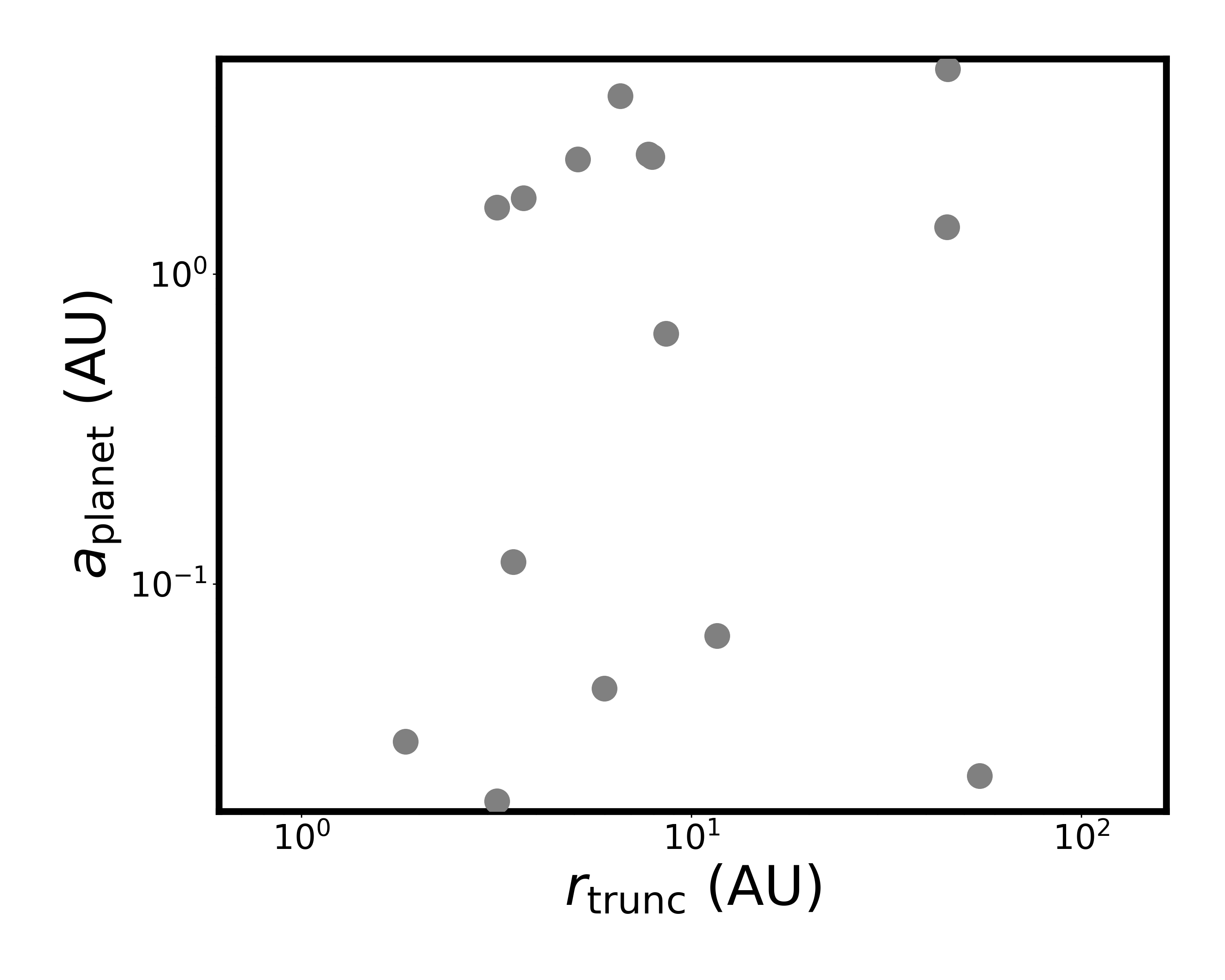}
    \includegraphics[width=0.48\textwidth]{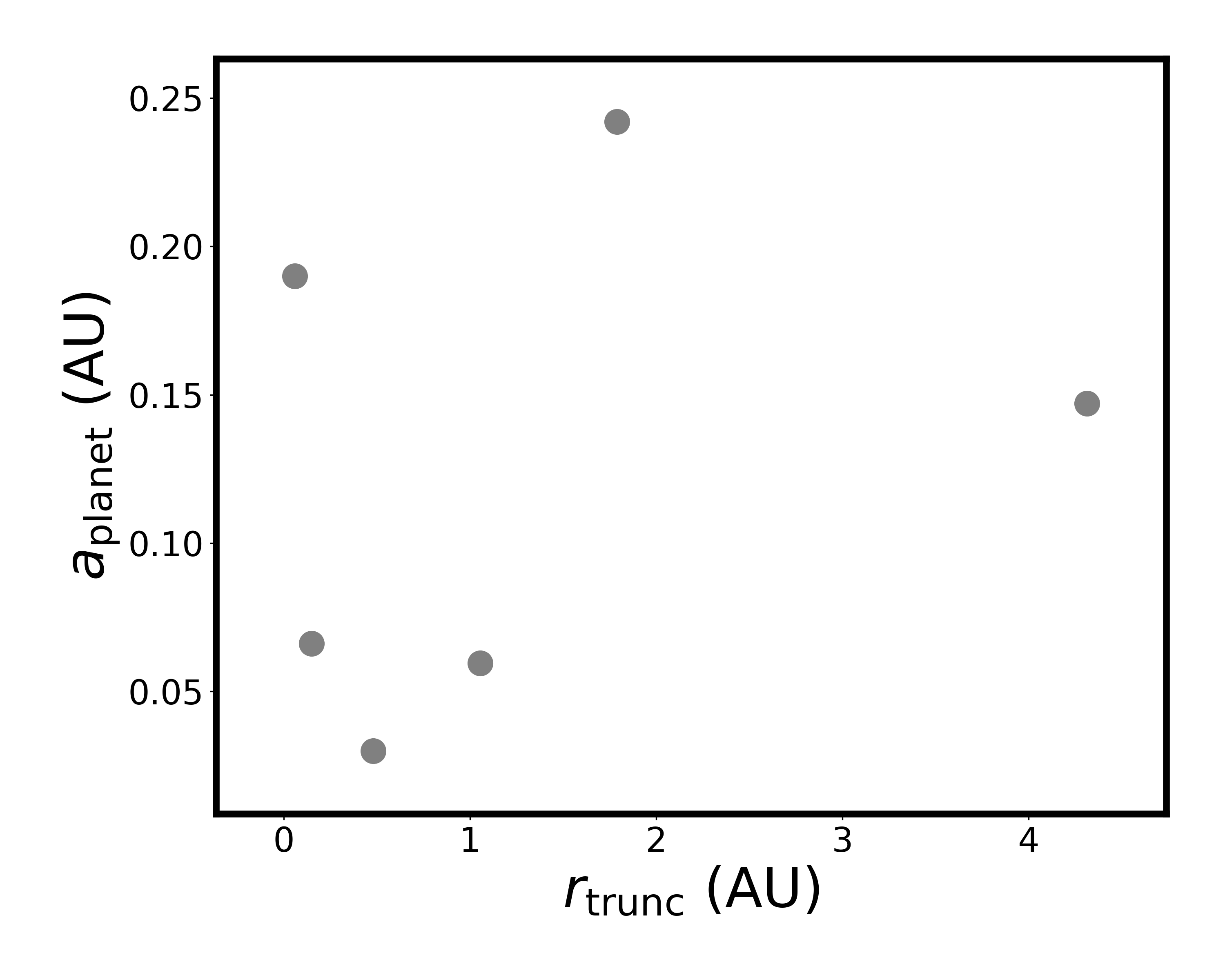}
    \caption{$r_t$ vs $a_\mathrm{planet}$ for gas giants with stable
    snow lines (upper panel) and terrestrial planets with unstable snow
    lines (lower panel). Only systems with measured eccentricity are
    shown.}
    \label{fig:trap_vs_observed_stable_giants_unstable_terrestrial}
\end{figure}

\begin{figure*}[h]
    \centering
    \includegraphics[width=0.48\textwidth]{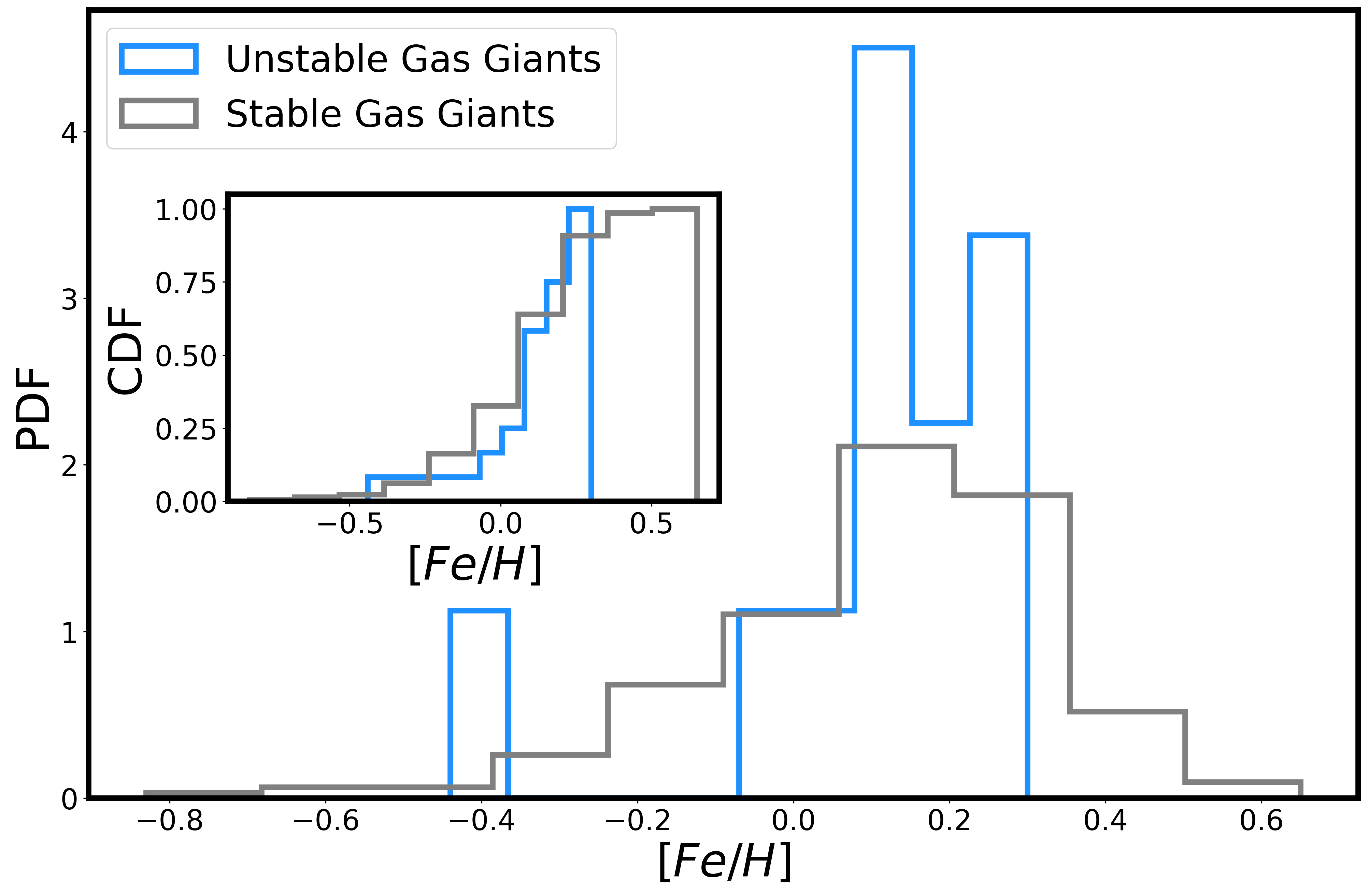}
    \hfill
    \includegraphics[width=0.48\textwidth]{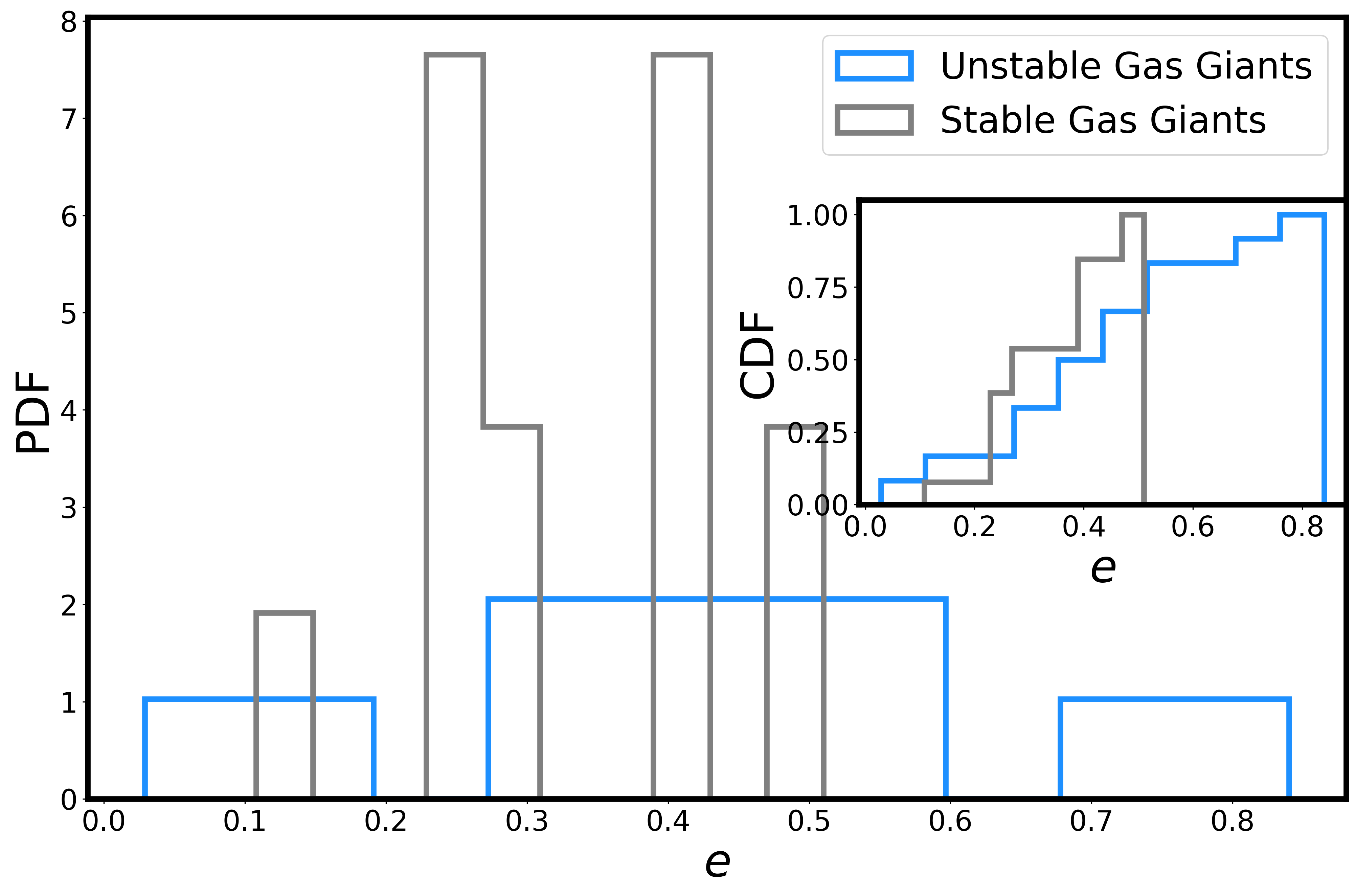}
    \caption{Comparison of metallicity (left) and eccentricity (right)
    distributions between gas giants with unstable (blue) and stable
    (grey) snow lines. Insets show the cumulative distribution function.
    Kolmogorov--Smirnov p-values: 0.64 (metallicity), 0.37
    (eccentricity), consistent with the two samples being drawn from the
    same parent distribution.}
    \label{fig:feh_e_hists}
\end{figure*}

\FloatBarrier  

\section{Sample parameters}
\label{app:table}

\begin{table*}[h]
\centering
\caption{Stellar and planetary parameters for the sample of unstable gas
giants. Systems marked with $\dagger$ are excluded from the linear-fit
analysis (see Sect.~\ref{sec:results}).}
\label{tab:stellar_params}
\small
\resizebox{0.8\textwidth}{!}{%
\begin{tabular}{lccccc}
\hline\hline
System & $m_1\,(M_\odot)$ & $m_2\,(M_\odot)$ & $e_b$ &
$a_b\,(\mathrm{AU})$ & $a_\mathrm{planet}\,(\mathrm{AU})$ \\
\hline
$\nu$~Oct$^\dagger$     & $1.57\pm0.06$ & $0.57\pm0.01$   & $0.24\pm0.0003$ & $2.6$          & $1.25$          \\
Kepler~693              & $0.80\pm0.035$& $0.145\pm0.05$  & $0.48\pm0.085$  & $2.9\pm0.6$    & $0.11\pm1.5$    \\
Kepler~420              & $0.99\pm0.05$ & $0.7\pm0.07$    & $0.31\pm0.29$   & $5.3\pm1.3$    & $0.382\pm0.006$ \\
HIP~90988$^\dagger$     & $1.30\pm0.08$ & $0.1\pm0.006$   & $0.29$          & $7.48$         & $1.36$          \\
HD~72892                & $1.02\pm0.05$ & $0.073\pm0.04$  & $0.38\pm0.003$  & $8.124$        & $0.228\pm0.008$ \\
TOI~1736                & $1.08\pm0.04$ & $0.78$          & ---             & $8.2$          & $1.35\pm0.017$  \\
HD~28192$^\dagger$      & $1.08$        & $0.09\pm0.008$  & $0.029$         & $8.5$          & $0.118$         \\
HD~59686$^\dagger$      & $1.86\pm0.2$  & $0.53\pm0.001$  & $0.729\pm0.004$ & $13.56\pm0.16$ & $1.086$         \\
HD~87646                & $1.12\pm0.09$ & $0.74$          & $0.54$          & $19.5$         & $1.58$          \\
$\gamma$~Cephei         & $1.18$        & $0.36\pm0.022$  & $0.41\pm0.007$  & $19.56\pm0.18$ & $2.05$          \\
HD~196885               & $1.1$         & $0.45\pm0.01$   & $0.42\pm0.02$   & $19.78\pm0.86$ & $2.383$         \\
HD~145934$^\dagger$     & $2.05$        & $0.085\pm0.042$ & $0.191\pm0.053$ & $21.774$       & $4.89$          \\
HD~5608                 & $1.29\pm0.23$ & $0.12\pm0.008$  & $0.53\pm0.22$   & $30.5$         & $1.91$          \\
HD~8673                 & $1.56\pm0.1$  & $0.38\pm0.07$   & $0.5\pm0.25$    & $35$           & $3.02$          \\
HIP~107773              & $2.39\pm0.27$ & $0.63\pm0.04$   & ---             & $40$           & $0.72\pm0.03$   \\
TOI~4633                & $1.1\pm0.06$  & $1.05\pm0.06$   & $0.91\pm0.03$   & $48.6\pm3.95$  & $0.847$         \\
$\gamma$~Leo$^\dagger$  & $1.41\pm0.14$ & $1.55\pm0.08$   & $0.84$          & $163$          & $1.19$          \\
\hline
\end{tabular}}
\end{table*}

\end{appendix}

\end{document}